# Infrared observation of OC-$C_2H_2$, OC-$(C_2H_2)_2$ and their isotopologues

A.J. Barclay,[a] A. Mohandesi,[a] K.H. Michaelian,[b] A.R.W. McKellar,[c] and N. Moazzen-Ahmadi[a]

[a] *Department of Physics and Astronomy, University of Calgary, 2500 University Drive North West, Calgary, Alberta T2N 1N4, Canada*

[b] *Natural Resources Canada, CanmetENERGY, 1 Oil Patch Drive, Suite A202, Devon, Alberta T9G 1A8, Canada*

[c] *National Research Council of Canada, Ottawa, Ontario K1A 0R6, Canada*

**Abstract**

The fundamental band for the OC-$C_2H_2$ dimer and two combination bands involving the intermolecular bending modes $\nu_9$ and $\nu_8$ in the carbon monoxide CO stretch region are re-examined. Spectra are obtained using a pulsed supersonic slit jet expansion probed with a mode-hop free tunable infrared quantum cascade laser. Analogous bands for OC-$C_2D_2$ and the fundamental for OC–DCCH as an impurity are also observed and analysed. A much weaker band in the same spectral region is assigned to a new mixed trimer, CO-$(C_2H_2)_2$. The trimer band is composed uniquely of *a*-type transitions, establishing that the CO monomer is nearly aligned with the *a*-inertial axis. The observed rotational constants agree well with *ab initio* calculations and a small inertial defect value indicates that the trimer is planar. The structure is a compromise between the T-shaped structure of free acetylene dimer and the linear geometry of free OC-$C_2H_2$. A similar band for the fully deuterated isotopologue CO-$(C_2D_2)_2$ confirms our assignment.

Address for correspondence:	Prof. N. Moazzen-Ahmadi,
	Department of Physics and Astronomy,
	University of Calgary,
	2500 University Drive North West,
	Calgary, Alberta T2N 1N4,
	Canada.

* Corresponding author. Tel: 1-403-220-5394

## 1. Introduction

There are a fair number of studies of the binary complex of carbon monoxide with acetylene. Initial studies were made through observation of 3 μm infrared spectra of $O^{12}C$-HCCH and $O^{13}C$-HCCH which showed that the dimer is linear, and that the carbon atom of the CO monomer lies between the center of mass of CO and that of the entire complex [1,2]. The infrared rotational constants quickly led to detection of pure rotational spectra for several isotopologues [3,4,5]. This body of data confirmed that $OC-C_2H_2$ is linear, with acetylene acting as the proton donor in forming the intermolecular bond. A second 3 μm study of OC-HCCH was made by Beck et al. [6] with higher resolution and sensitivity. They observed extensive perturbations which were attributed to high density of states at 3300 cm$^{-1}$. These authors then argued that as density of states increases the probability of perturbation with nearby levels grows until at some point the states are so badly perturbed that the spectrum will appear to be random and chaotic. This claim was later proven to be incorrect by observation of infrared spectra of systems with higher density of states, such as deuterated species, which hardly show any perturbation [7]. It was then concluded that the extensive perturbations in the normal acetylene complex are due to a simple coincidence of energy levels, rather than any anomaly in coupling strengths or densities of states.

Later infrared studies were made in the 4.7 μm CO stretching region [8,9]. In contrast to the extensive perturbations observed in the 3 μm region, spectra in the CO stretching fundamental showed only a single small perturbation. More recently, spectroscopic studies of OC-HCCH were extended by Rivera-Rivera et al. [10] to include the intermolecular bending modes $\nu_9$ and $\nu_8$ in the carbon monoxide CO stretch region. They used quantum cascade lasers to measure $\nu_3$, $\nu_3+\nu_9$, $\nu_3+\nu_9-\nu_9$, and $\nu_3+\nu_8$ (for consistency, we retain here their vibrational mode notation for OC-HCCH: $\nu_3$ is the C-O stretch ($\approx$2149 cm$^{-1}$), $\nu_5$ is the van der Waals stretch ($\approx$50 cm$^{-1}$), and $\nu_8$ and $\nu_9$ are the van der Waals bends ($\approx$70 and 20 cm$^{-1}$)). Although Table 1 in their work [10] gives experimentally determined spectroscopic constants for the four vibrational states involved, Rivera-Rivera et al. do not provide a full analysis of $\nu_3+\nu_8$, only briefly mentioning that a preliminary analysis indicated significant perturbations in the upper state. We show below that, at least for *J*-values up to 11, this is not the case.

There are virtually no high resolution spectroscopic data on larger clusters involving CO and other small linear molecules. One exception is the recent detection of CO-(CO$_2$)$_2$ [11] where a quasi T-shaped trimer structure is observed. This structure effectively combines the structures of the CO$_2$-CO$_2$ dimer (planar slipped parallel) and the CO$_2$-CO dimer (T-shaped, with CO as the stem of the T, and adjacent C atoms).

In this paper, we re-examine the $\nu_3$, $\nu_3+\nu_9$, and $\nu_3+\nu_8$ bands of OC-HCCH, and provide new data for the same three bands of OC-DCCD and the fundamental of OC-DCCH. We show that, although the bands for OC-HCCH exhibit small perturbations, they can be analysed to experimental accuracy using effective parameters. This is not possible for OC-DCCD because the upper state of $\nu_3+\nu_8$ shows extensive perturbations for both *e* and *f* parity states, indicating strong coupling with a "dark" vibrational state, and resulting in a several perturbation-allowed lines which are clearly visible.

Thanks to the high signal-to-noise ratio of our spectra, we were also able to observe a much weaker *a*-type band in the CO stretching fundamental region. A ro-vibrational analysis of this band indicates that it is due to a planar isomer of the mixed trimer OC-(C$_2$H$_2$)$_2$. This conclusion is further confirmed by the presence of an analogous band for OC-(C$_2$D$_2$)$_2$ in the same spectral region and by *ab initio* calculations at MP2/cc-pvtz level. This structure, which corresponds to an asymmetric rotor with $C_s$ point group symmetry, preserves fairly closely the T-shaped geometry of acetylene dimer and the linear geometry of OC-HCCH complex. In many respects, this structure parallels previously reported planar isomers of OCS-$(C_2H_2)_2$ [12] and $N_2O$-$(C_2H_2)_2$ [13].

## 2. Observed spectra

The spectra were recorded at the University of Calgary as described previously [14-16], using a pulsed supersonic slit jet apparatus and a Daylight Solutions quantum cascade laser. A typical gas mixture consisted of about 0.2 % CO and 0.2 % $C_2H_2$ in helium carrier gas, and a jet backing pressure of 9 atmospheres were used. Spectral assignment and simulation were made using PGOPHER [17].

### 2.1 OC-HCCH and OC-DCCD results

The experimental spectra for the $\nu_3$ stretching fundamental (a $\Sigma$ - $\Sigma$ ro-vibrational band) of OC-HCCH (top trace) and OC-DCCD (third trace) are illustrated in Fig. 1. The simulated spectra (second and fourth trace) have a rotational temperature of 2.2 K. Due to the relatively low rotational temperature, we were able to observe and include only 24 lines for OC-HCCH and 25 for OC-DCCD in the final analyses.

Figure 2 shows a segment of the experimental spectrum for the $\nu_3+\nu_9$ combination band of OC-HCCH (top trace), a linear molecule $\Pi$ - $\Sigma$ band with a characteristic $Q$-branch. We found no evidence for any sizeable perturbations, and the band can therefore be well simulated using effective parameters. The simulated spectrum for a rotational temperature of 2.2 K is shown in the second trace in Fig. 2. The corresponding experimental and simulated spectra for OC-DCCD are shown in the third and fourth traces, respectively. We assigned 35 lines for OC-HCCH and the same number for OC-DCCD.

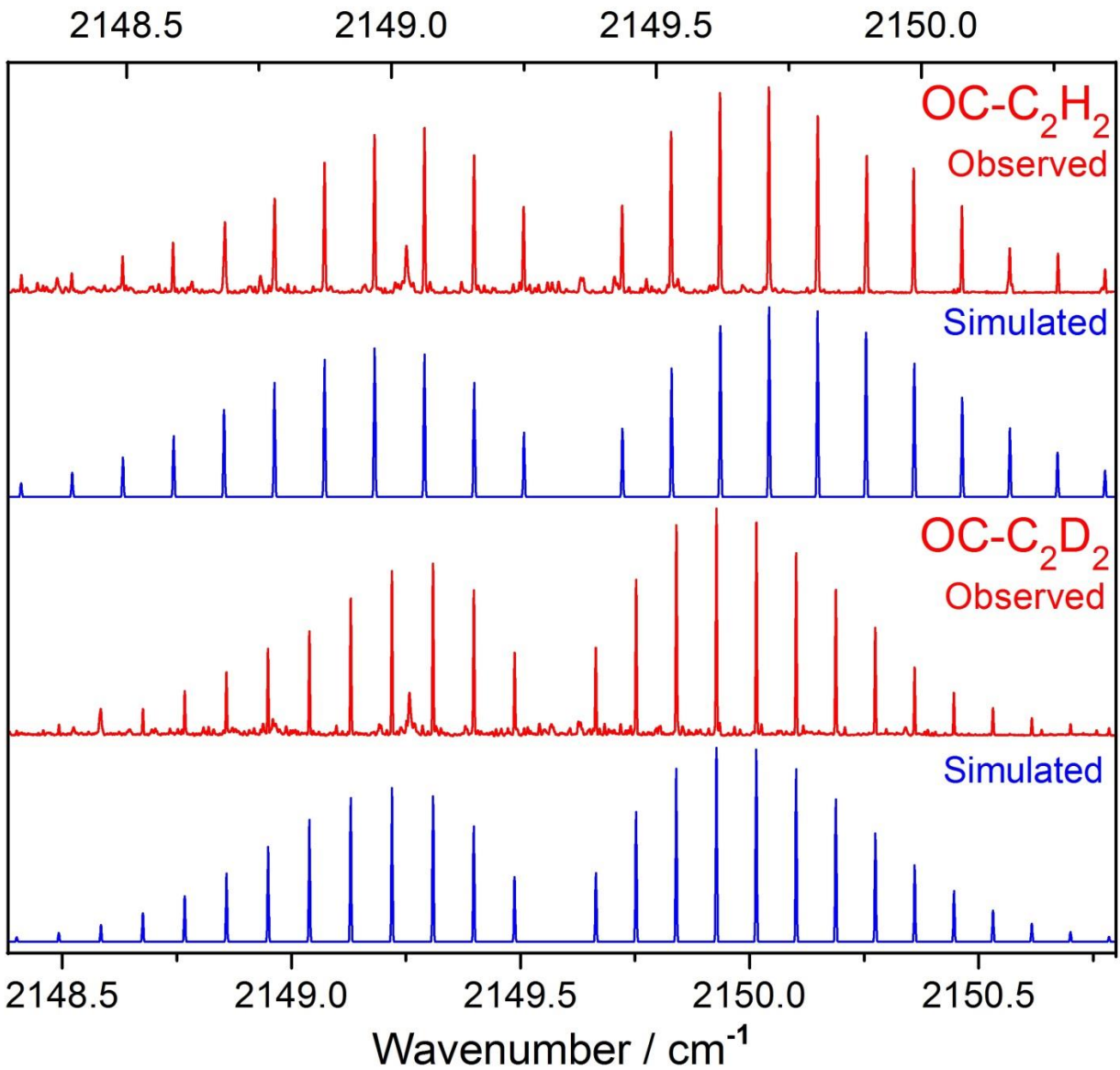

Figure 1: Observed (first and third trace) spectra of the $\nu_3$ fundamental band for OC-HCCH and OC-DCCD. The simulated spectra (second and fourth trace) were made at a rotational temperature of 2.2 K

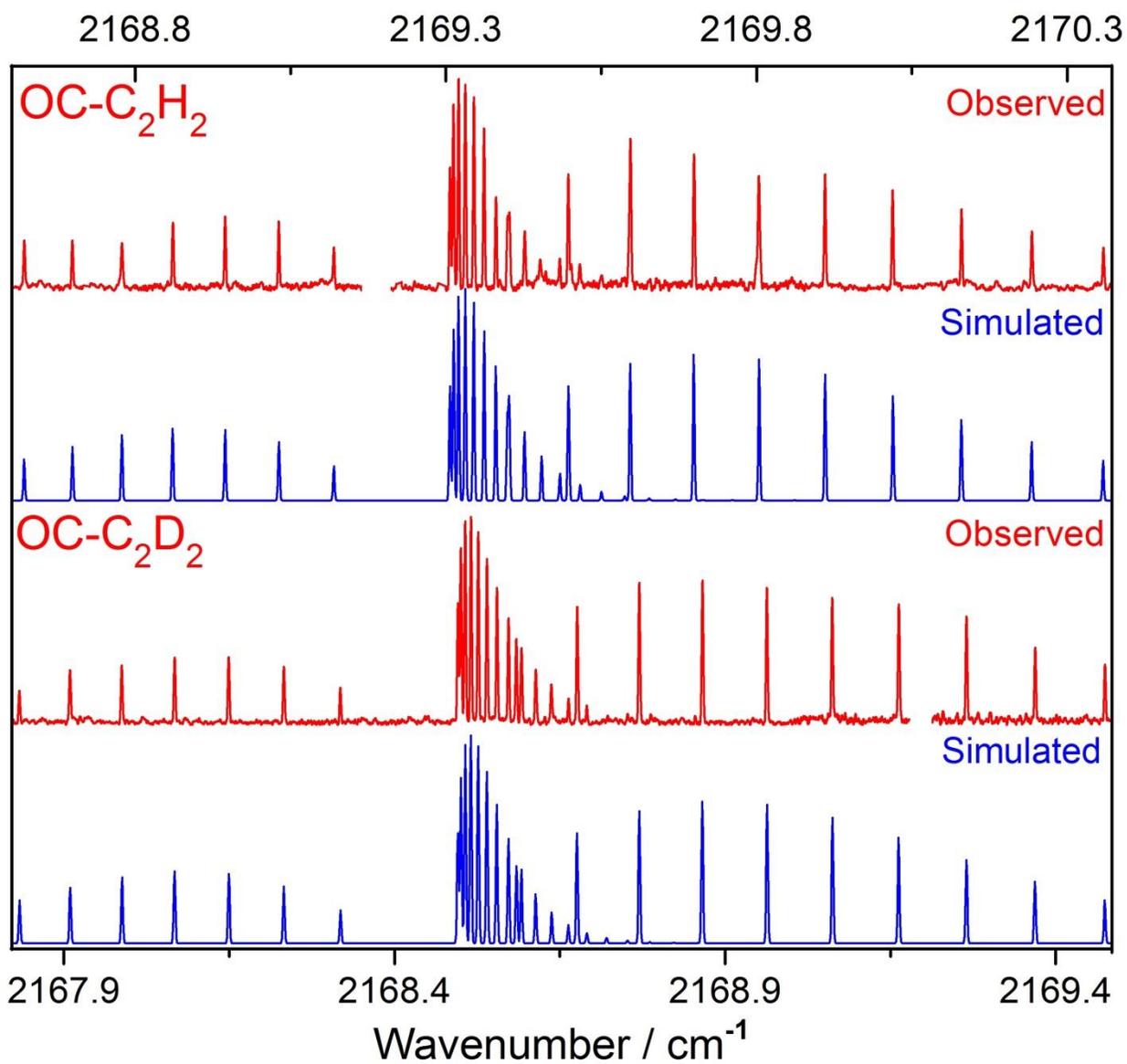

Figure 2: Observed (first and third trace) spectra of the $\nu_3+\nu_9$ combination band for OC-HCCH and OC-DCCD. The simulated spectra (second and fourth trace) were made at a rotational temperature of 2.2 K. The blank regions in the experimental spectra are CO monomer lines.

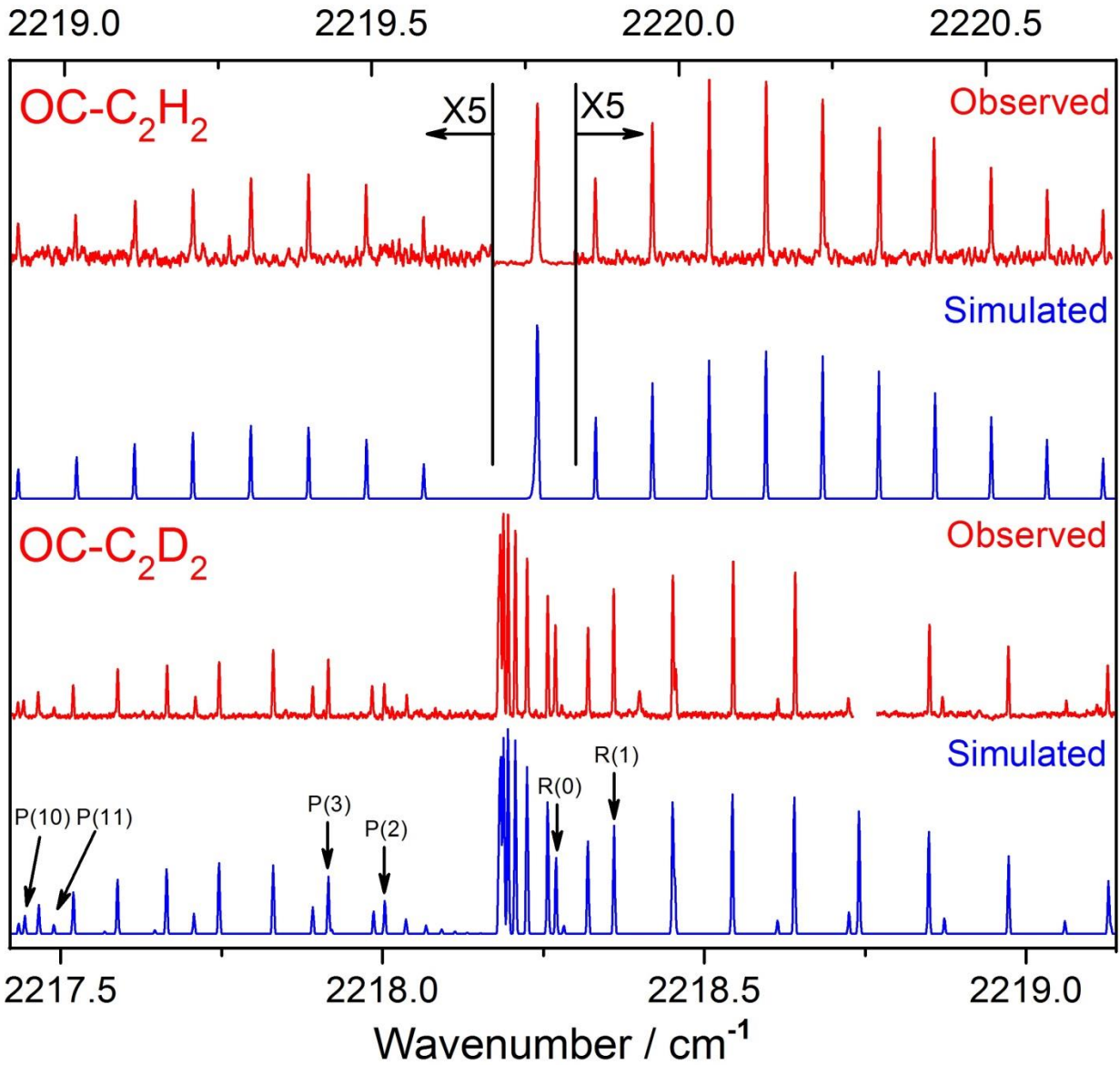

Figure 3: Observed (first and third trace) spectra of the $\nu_3+\nu_8$ combination band for OC-HCCH and OC-DCCD. The simulated spectra (second and fourth trace) were made at a rotational temperature of 2.2 K. The blank region in the OC-DCCD experimental spectrum is a CO monomer line.

The combination band $\nu_3+\nu_8$ is also a $\Pi$ - $\Sigma$ band. A segment of this spectrum is shown in the top trace of Fig. 3. Here, the strong, unresolved and slightly red shaded feature at 2219.769 cm$^{-1}$ is the $Q$-branch. Perhaps the striking difference between this $Q$-branch and the fully resolved $Q$-branch of the $\nu_3+\nu_9$ band led Rivera-Rivera et al. [10] to conclude that $\nu_3+\nu_8$ is badly perturbed. However, we are able to analyse this band with no difficulty using effective parameters and including 22 lines from the $P$ and $R$ branches. Since our effective rotational temperature is significantly lower than theirs, it remains possible that the perturbations mentioned by Rivera-Rivera et al. [10] affected only transitions with $J$-values higher than about 11.

Spectroscopic constants for the $\nu_3$, $\nu_3+\nu_9$, and $\nu_3+\nu_8$ bands of OC-HCCH were retrieved by keeping the ground rotational constants fixed at their microwave values [5] and varying the rotational constants of the three excited vibrational states. A total of 81 frequencies were used in the analysis and 9 parameters were varied. The $l$-doubling parameter $q$ for $\nu_3+\nu_8$ was varied step-wise to reproduce the slightly red-shaded $Q$-branch in the top trace of Fig. 3. The rms error of the fit was 0.00052 cm$^{-1}$. The molecular parameters obtained are listed in Table 1 and detailed assignments are given in the supplementary material.

Table 1. Molecular parameters for OC-HCCH (in cm$^{-1}$).[a]

| | Ground state | Excited state $\nu_3$ fundamental | Excited state $\nu_3+\nu_9$ combination | Excited state $\nu_3+\nu_8$ combination |
|---|---|---|---|---|
| $\nu_0$ | | 2149.3430(2) | 2169.3039(2) | 2219.7713(2) |
| $B$ | 0.046611254[b] | 0.0464517(18) | 0.0478620(44) | 0.0465126(85) |
| $10^7\,D_J$ | 1.759[b] | 1.759[b] | 2.86(21) | 1.11(58) |
| $10^4\,q$ | | | −2.457(25) | −0.45[c] |

[a] Uncertainties (1σ) in parentheses are in units of the last quoted digit.

[b] Held fixed at the value determined from rotational spectrum [5].

[c] Varied step-wise to reproduce the slight red-shaded Q-branch. See Fig. 3

The $\nu_3+\nu_8$ combination band of OC-DCCD is strongly perturbed (see the third trace in Fig. 3). The clearest indication for this is in the *P*- and *R*-branches where the *P*-branch shows a rapid decrease in rotational spacing and eventual turn around at *P*(10) with a concomitant rapid increase in rotational spacing in the *R*-branch at higher *J* values. The rotational progression in the *Q*-branch is also very irregular. The indication is that the perturbing "dark" state lies at lower energy, and analysis of this band, including 38 transitions (with four perturbation allowed lines) reveals that the origin of the dark state is located about 2 cm$^{-1}$ below the $\nu_3+\nu_8$ band center. A total of 97 transitions for the four vibrational bands of OC-DCCD were used in the analysis and 17 parameters were varied. The fit had an rms error of 0.00076 cm$^{-1}$, about twice the estimated (relative) experimental accuracy of 0.0003 cm$^{-1}$. The main contribution to the excess rms came from transitions in the $\nu_3+\nu_8$ band. The molecular parameters obtained are listed in Table 2 and detailed assignments are given in the supplementary material.

Table 2. Molecular parameters for OC-DCCD (in cm$^{-1}$).[a]

|  | Ground state | Excited state $\nu_3$ fundamental | Excited state $\nu_3+\nu_9$ combination | Excited state $\nu_3+\nu_8$ combination | "Dark" state[c] |
|---|---|---|---|---|---|
| $\nu_0$ |  | 2149.5760(3) | 2168.4936(3) | 2218.1800(3) | 2215.956(3) |
| $B$ | 0.044342977[b] | 0.0441911(83) | 0.0453263(82) | 0.044779(13) | 0.064302(35) |
| $10^7 \, D_J$ | 1.438[b] | 1.28(44) | 2.24(47) | −16.92(94) |  |
| $10^4 \, q$ |  |  | −2.307(32) | −2.81(17) | −35.83(20) |

[a] Uncertainties (1σ) in parentheses are in units of the last quoted digit.

[b] Held fixed at the values determined from rotational spectrum [4].

[c] The interaction Hamiltonian connecting the dark state and $\nu_3+\nu_8$ is given by $H_{int} = \gamma_1 \left( J_+^2 l_-^2 + J_+^2 l_-^2 \right) + \gamma_2 J^2 + \gamma_3 J^4$. The values for the interaction parameters were determined to be $\gamma_1 = 3.4346(38) \times 10^{-3}$ cm$^{-1}$, $\gamma_2 = -7(3) \times 10^{-5}$ cm$^{-1}$, $\gamma_3 = -2.38(23) \times 10^{-6}$ cm$^{-1}$.

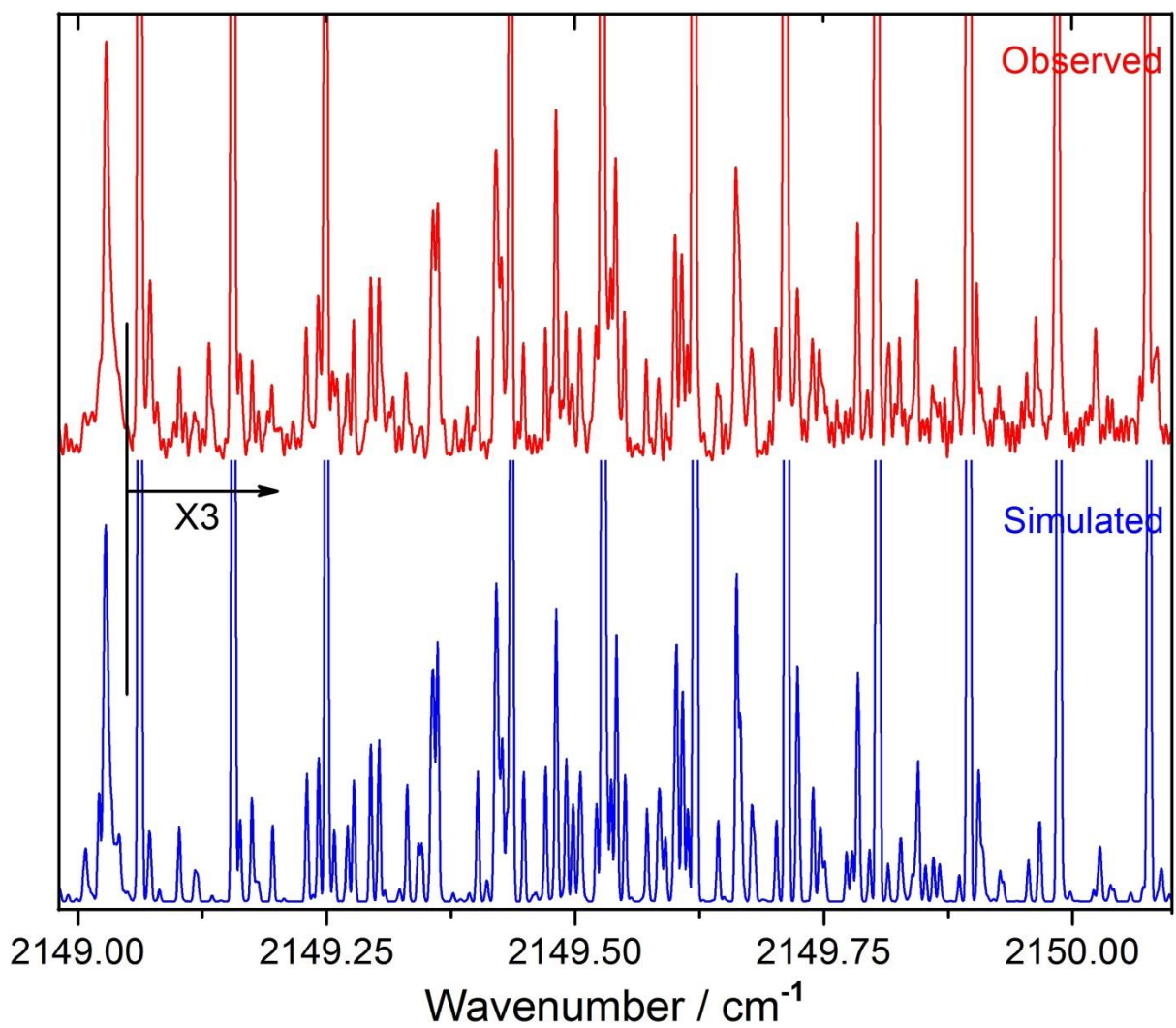

Figure 4: Observed (top trace) spectra of fundamental band for OC-$(C_2H_2)_2$. The simulated spectrum (bottom trace) was made at a rotational temperature of 2.2 K and a 0.003 cm$^{-1}$ Gaussian line width (FWHM).

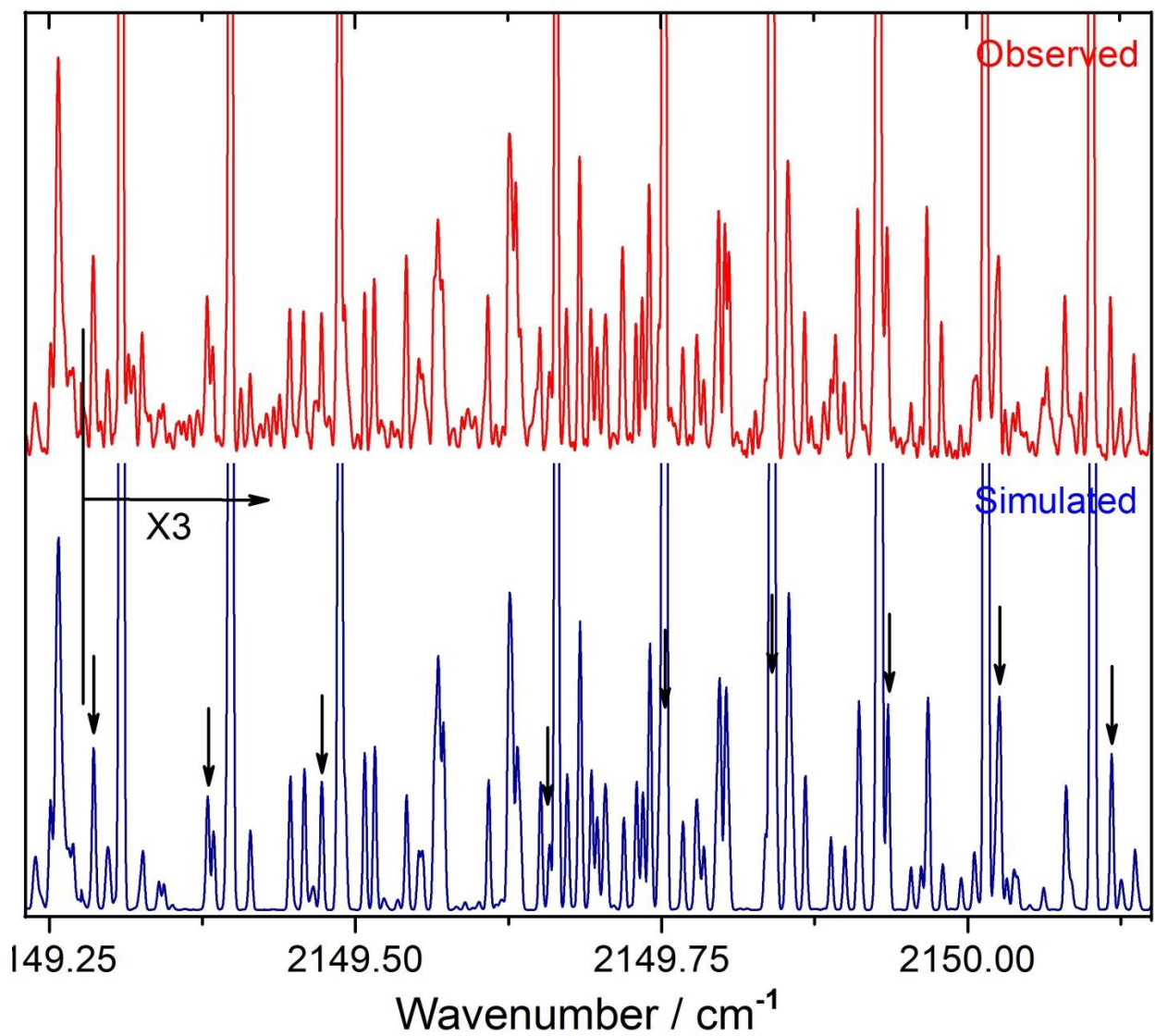

Figure 5: Observed (top trace) spectra of fundamental band for OC-$(C_2D_2)_2$. The simulated spectrum (bottom trace) was made at a rotational temperature of 2.2 K and a 0.003 cm$^{-1}$ Gaussian line width (FWHM). The lines marked by black arrows are the transitions in the CO stretching fundamental band of OC-DCCH.

2.2 OC-$(C_2H_2)_2$ and OC-$(C_2D_2)_2$ results

After accounting for the known dimer transitions in the fundamental bands, we were left with a large number of weak unassigned lines in the same spectral region for both OC-HCCH and OC-DCCD. This can be seen in the first and third traces in Fig. 1. The relatively strong feature between the rotational lines P(3) and P(4) of the $\nu_3$ fundamental, evident in both the OC-HCCH and OC-DCCD spectra, led us to conclude that this is the $Q$-branch of an $a$-type band of a complex containing CO and acetylene. We eventually assigned this band to a new mixed trimer, OC-$(C_2H_2)_2$. The top trace in Fig. 4 shows a close-up of the top trace in Fig. 1. The strongest features (off scale) are the rotational lines of OC-HCCH. The broad feature at about 2149.04 cm$^{-1}$ is the $a$-type $Q$-branch of the mixed trimer and the lines to higher frequency, whose intensities are magnified by a factor of 3, constitute the $R$-branch. The lower trace is the simulated spectrum with a rotational temperature of 2.2 K and a Gaussian line width of 0.003 cm$^{-1}$. The simulation indicates that this band consists uniquely of $a$-type transitions, and hence the $a$-inertial axis of the trimer must approximately coincide with the carbon monoxide C-O axis. Analogous spectra for OC-$(C_2D_2)_2$ shown in Fig. 5 lead to the same conclusion.

Ultimately we assigned a total of 48 observed lines in terms of 52 transitions of CO-$(C_2H_2)_2$, with values of $J$ and $K_a$ ranging up to 11 and 4, respectively. These were used in a least-squares fit to determine the set of 7 parameters listed in Table 3. The average error in the fit was 0.00048 cm$^{-1}$. This fit was obtained without introducing any centrifugal distortion parameters. The assignments are given as Supplementary data.

Similarly, we assigned 42 frequencies in terms of 49 transitions of OC-$(C_2D_2)_2$ with values of $J$ and $K_a$ ranging up to 10 and 4, respectively. The least-squares fit gave an average error of 0.00045 cm$^{-1}$ with a set of 7 parameters which are listed in Table 3. Again, no centrifugal distortion parameters were required. The assignments are given as Supplementary data.

Table 3. Molecular parameters for $CO-(C_2H_2)_2$ and $CO-(C_2D_2)_2$. [a]

|  | This work | | MP2 /cc-pvtz | |
| --- | --- | --- | --- | --- |
|  | $CO-(C_2H_2)_2$ | $CO-(C_2D_2)_2$ | $CO-(C_2H_2)_2$ | $CO-(C_2D_2)_2$ |
| $\nu_0$ / cm$^{-1}$ | 2149.0241(1) | 2149.2538(1) | | |
| $A'$ / MHz | 2668.8(64) | 2536.1(69) | | |
| $B'$ / MHz | 1405.04(57) | 1315.34(52) | | |
| $C'$ / MHz | 918.49(76) | 862.12(31) | | |
| $A''$ / MHz | 2661.4(64) | 2528.4(69) | 2715 | 2552 |
| $B''$ / MHz | 1410.23(65) | 1320.85(67) | 1446 | 1351 |
| $C''$ / MHz | 920.61(95) | 863.23(33) | 944 | 883 |
| $\Delta''$ / amu Å$^2$ | 0.702 | 2.95 | | |

[a] Uncertainties (1σ) in parentheses are in units of the last quoted digit.

## 3. Discussion and conclusions

The lines marked by arrows in Fig. 5 are transitions in the CO stretching fundamental of OC-DCCH due to singly deuterated acetylene existing as an impurity in the $C_2D_2$ sample. We assigned 17 transitions from this band. The analysis was made with the rotational parameters fixed at their microwave values [3]. The resulting parameters for the excited state are: $\nu_0 = 2149.5660(1)$ cm$^{-1}$, $B' = 0.0463423(57)$ cm$^{-1}$, $D' = 9.5(45) \times 10^{-8}$ cm$^{-1}$. The rms error of the fit was 0.00028 cm$^{-1}$. The assignments are given as Supplementary data.

Several *ab initio* calculations have been reported for carbon monoxide–acetylene [18,19,20,21]. All verify that OC–HCCH is linear, but the reported intermolecular frequencies are not very consistent, with a large spread between the various theoretical values and significant differences from observed frequencies (see Table 1 of Ref. [10]). Perhaps the most reliable prediction for the intermolecular frequencies to date is that of Rivera-Rivera et al. [10], who used experimental rovibrational data involving $\nu_9$ together with other spectroscopic information to generate a four-dimensional morphed potential. This potential was then used to further characterize the structure and frequencies for other intermolecular modes. Their predictions for OC-HCCH were: $D_e = 360(4)$ cm$^{-1}$, $D_0 = 232(4)$ cm$^{-1}$, $\nu_5 = 53.5(4)$ cm$^{-1}$ (the hydrogen bond stretching frequency), and $\nu_8 = 73.7(2)$ cm$^{-1}$ (the higher frequency intermolecular bend).

The experimental intermolecular frequencies from Ref. [10] and from the present work are listed in Table 4. Because the intermolecular and intramolecular modes do not couple strongly, we expect the intermolecular frequencies for $\nu_9$ in the ground vibrational state (20.48361(20) cm$^{-1}$) and in the excited $\nu_3$ state (19.961(1) cm$^{-1}$) to differ only slightly, as is indeed the case. So it is reasonable to assume that other excited state intermolecular frequencies listed in Table 4 are also close to their values in the ground vibrational state. We note that this assertion is valid for many other binary clusters [22].

Table 4. Observed intermolecular frequencies (in cm$^{-1}$) for OC-HCCH and OC-DCCD in the excited $\nu_3$ state

|  | OC-HCCH | | OC-DCCD |
|---|---|---|---|
|  | This work | Ref. [10] [a] | This work |
| $\nu_9$ | 19.961(1) | 19.961 | 18.918(1) |
| $\nu_8$ | 70.428(1) | 70.429 | 68.604(1) |

[a] The observed intermolecular frequency for $\nu_9$ in the ground vibrational state is 20.48361(20) cm$^{-1}$.

As mentioned previously, the upper state of the $\nu_3+\nu_8$ combination band of OC-DCCD is strongly perturbed, and our analysis indicates that the origin of the perturbing state is located at 2216 cm$^{-1}$. Our best guess for a perturbing state with correct symmetry is $\nu_3+\nu_5+\nu_9$. This assignment would result in a value of $\nu_5 = 48(2)$ cm$^{-1}$ for the hydrogen bond stretching frequency, which seems reasonable for OC-DCCD since the predicted value for OC-HCCH in Ref. [10] is $\nu_5 = 53.5(4)$ cm$^{-1}$.

This (or indeed any other) vibrational assignment of the dark state does not explain the anomalously large values of $B$ and $q$ reported in Table 4 for this state. Explanation of this anomaly may have to wait for the availability of higher level *ab initio* calculations which fully explore the complete intermolecular potential surfaces for CO-C$_2$H$_2$ and CO-C$_2$D$_2$. We are aware that such a surface is under construction and to be published in the near future [23]. Such calculations may also help to explain why the O-bonded isomer, CO-HCCH, has not been experimentally detected.

We are very confident that the spectra observed in Fig. 4 belong to the planar mixed trimer CO-$(C_2H_2)_2$. This conclusion is supported by the small inertial defect of 0.72 amu Å$^2$, by the good fit to the observed spectra, by consistent isotope shifts, and by the observed *a*-type selection rules. For further verification, we made molecular structure calculations using the Gaussian-09 program. The calculations were carried out at the second order level of Moller-Plesset perturbation theory (MP2) in the same manner as those reported in Ref. [11] for CO-$(CO_2)_2$. The resulting rotational parameters for OC-HCCH and OC-DCCD are listed in Table 4. They compare favorably with their experimental counterparts and help to confirm a planar structure. This structure and its inertial axes are illustrated in Fig. 6. As can be seen, the CO monomer lies close to the *a*-inertial axis, consistent with the observed *a*-type structure of the band.

Although analysis of our results confirms the geometry shown in Fig. 6 and gives structural information for the trimer, they are not sufficient for determination of a unique experimental structure, which would require six or more rotational parameters. Here, we only have four independent rotational parameters. Finally, it should be noted that we observe no evidence for tunneling splitting in the spectra of the trimer and conclude that the large amplitude tunneling motion that exists in the free acetylene dimer is quenched by presence of the CO monomer. This is also the case for OCS-$(C_2H_2)_2$ [12] and $N_2O$-$(C_2H_2)_2$ [13].

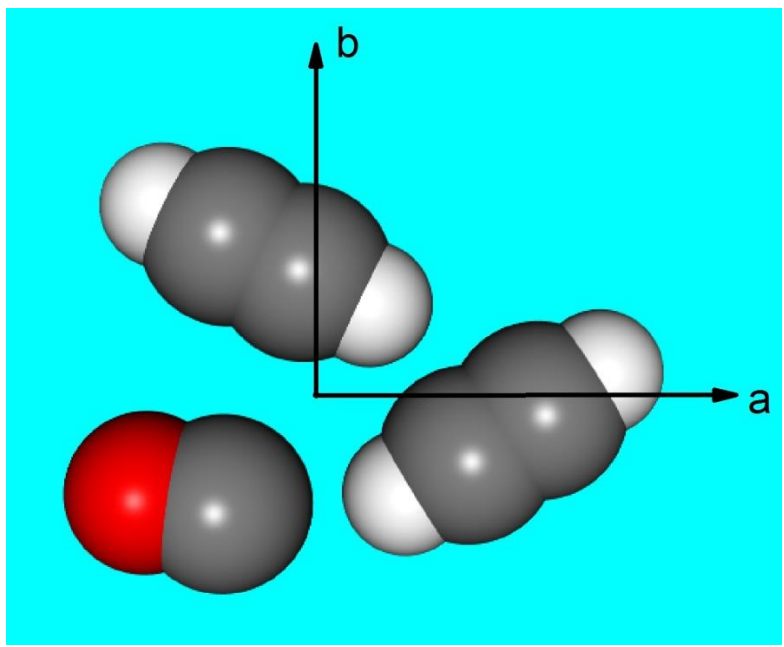

Figure 6: The mixed CO-$(C_2H_2)_2$ trimer and its inertial axes.

In conclusion, mid-infrared studies of a new planar isomer of the mixed trimer CO-$(C_2H_2)_2$, are reported in the carbon monoxide CO stretch region using a mode-hop free tunable infrared quantum cascade laser. The trimer band is composed uniquely of *a*-type transitions, establishing that the CO monomer is nearly aligned with the *a*-inertial axis. The observed rotational constants agree well with *ab initio* calculations and a small value of the inertial defect indicates that the trimer is planar. This structure combines the T-shaped structure of free acetylene dimer and the linear geometry of free OC-$C_2H_2$. Analysis of a similar band for the fully deuterated isotopologue CO-$(C_2D_2)_2$ confirms our assignment. In addition, the $\nu_3$ fundamental band of the linear dimer OC-$C_2H_2$, and two combination bands involving the intermolecular modes $\nu_9$ and $\nu_8$, are re-examined. New spectra are observed for OC-$C_2D_2$, and its intermolecular bending frequencies are determined by observation of combination bands. Finally, the fundamental band OC–DCCH, observed as an impurity in the OC-$C_2D_2$ spectrum, is observed and analysed.

## Acknowledgements

The financial support of the Natural Sciences and Engineering Research Council of Canada is gratefully acknowledged. We also thank K. Esteki and J. Norooz Oliaee for help with the experiments.

## Appendix A. Supplementary data

Supplementary data associated with this article can be found below.

Supplementary Material for:
Infrared observation of OC-C2H2, OC-(C2H2)2 and their isotopologues

A.J. Barclay, A. Mohandesi, K.H. Michaelian, A.R.W. McKellar, N. Moazzen-Ahmadi

Table A-1. Observed and calculated transitions in the C-O
stretching fundamental band of OC-HCCH dimer (units of 1/cm).

```
**************************************************
  J' symm   J" symm    Obs.       Calc.     Obs-Calc
**************************************************
  10   e    11   e    2148.3016  2148.3010   0.0007
   9   e    10   e    2148.3973  2148.3971   0.0002
   8   e     9   e    2148.4928  2148.4930  -0.0002
   6   e     7   e    2148.6857  2148.6840   0.0017
   5   e     6   e    2148.7796  2148.7791   0.0006
   4   e     5   e    2148.8738  2148.8738   0.0000
   3   e     4   e    2148.9682  2148.9683  -0.0000
   2   e     3   e    2149.0622  2149.0624  -0.0002
   1   e     2   e    2149.1561  2149.1563  -0.0002
   0   e     1   e    2149.2495  2149.2498  -0.0003
   1   e     0   e    2149.4354  2149.4359  -0.0005
   2   e     1   e    2149.5279  2149.5285  -0.0006
   3   e     2   e    2149.6202  2149.6208  -0.0006
   4   e     3   e    2149.7123  2149.7127  -0.0004
   5   e     4   e    2149.8045  2149.8043   0.0003
   6   e     5   e    2149.8969  2149.8955   0.0014
   7   e     6   e    2149.9856  2149.9864  -0.0008
   8   e     7   e    2150.0765  2150.0770  -0.0004
   9   e     8   e    2150.1669  2150.1672  -0.0003
  10   e     9   e    2150.2579  2150.2570   0.0009
  11   e    10   e    2150.3463  2150.3465  -0.0001
  12   e    11   e    2150.4349  2150.4356  -0.0007
  13   e    12   e    2150.5241  2150.5243  -0.0003
  14   e    13   e    2150.6125  2150.6127  -0.0002
**************************************************
```

Table A-2. Observed and calculated transitions in nu3+nu9 combination band of OC-HCCH dimer in the region of C-O stretching fundamental (units of 1/cm).

```
**************************************************
  J' symm  J" symm    Obs.       Calc.     Obs-Calc
**************************************************
  10   e    11   e   2168.4023  2168.4021   0.0002
   9   e    10   e   2168.4731  2168.4730   0.0001
   8   e     9   e   2168.5460  2168.5460   0.0000
   7   e     8   e   2168.6213  2168.6213   0.0000
   6   e     7   e   2168.6987  2168.6987  -0.0001
   5   e     6   e   2168.7785  2168.7784   0.0001
   4   e     5   e   2168.8605  2168.8603   0.0002
   3   e     4   e   2168.9446  2168.9445   0.0000
   2   e     3   e   2169.0309  2169.0310  -0.0000
   1   e     2   e   2169.1196  2169.1197  -0.0001
   1   f     1   e   2169.3066  2169.3066  -0.0000
   2   f     2   e   2169.3120  2169.3121  -0.0001
   3   f     3   e   2169.3203  2169.3203   0.0000
   4   f     4   e   2169.3311  2169.3313  -0.0002
   5   f     5   e   2169.3450  2169.3450   0.0000
   6   f     6   e   2169.3613  2169.3614  -0.0000
   7   f     7   e   2169.3805  2169.3804   0.0001
   1   e     0   e   2169.3997  2169.3993   0.0003
   8   f     8   e   2169.4019  2169.4022  -0.0002
   9   f     9   e   2169.4268  2169.4266   0.0002
  11   f    11   e   2169.4833  2169.4833   0.0000
   2   e     1   e   2169.4970  2169.4971  -0.0001
  12   f    12   e   2169.5154  2169.5155  -0.0000
   3   e     2   e   2169.5968  2169.5970  -0.0002
   4   e     3   e   2169.6993  2169.6992   0.0001
   5   e     4   e   2169.8036  2169.8036   0.0000
   6   e     5   e   2169.9101  2169.9102  -0.0001
   7   e     6   e   2170.0189  2170.0190  -0.0001
   8   e     7   e   2170.1298  2170.1299  -0.0001
   9   e     8   e   2170.2431  2170.2430   0.0001
```

| | | | | | | |
|---|---|---|---|---|---|---|
| 10 | e | 9 | e | 2170.3583 | 2170.3581 | 0.0002 |
| 11 | e | 10 | e | 2170.4747 | 2170.4753 | -0.0006 |
| 12 | e | 11 | e | 2170.5950 | 2170.5946 | 0.0005 |
| 13 | e | 12 | e | 2170.7157 | 2170.7158 | -0.0002 |
| 14 | e | 13 | e | 2170.8387 | 2170.8390 | -0.0003 |
| 15 | e | 14 | e | 2170.9644 | 2170.9642 | 0.0002 |

**************************************************

Table A-3. Observed and calculated transitions in nu3+nu8 combination band of OC-HCCH dimer in the region of C-O stretching fundamental (units of 1/cm).

```
*********************************************
  J' symm  J" symm    Obs.       Calc.     Obs-Calc
*********************************************
  10  e    11   e    2218.8297  2218.8294   0.0003
   9  e    10   e    2218.9244  2218.9244   0.0000
   8  e     9   e    2219.0178  2219.0193  -0.0015
   7  e     8   e    2219.1149  2219.1140   0.0009
   6  e     7   e    2219.2091  2219.2085   0.0006
   5  e     6   e    2219.3034  2219.3029   0.0005
   4  e     5   e    2219.3971  2219.3970   0.0001
   3  e     4   e    2219.4907  2219.4909  -0.0002
   2  e     3   e    2219.5842  2219.5846  -0.0004
   1  e     2   e    2219.8639  2219.8643  -0.0003
   1  e     1   e    2219.9568  2219.9570  -0.0002
   2  e     2   e    2220.0495  2220.0495  -0.0000
   3  e     3   e    2220.1421  2220.1417   0.0003
   4  e     4   e    2220.2341  2220.2337   0.0003
   5  e     5   e    2220.3266  2220.3255   0.0011
   6  e     6   e    2220.4153  2220.4170  -0.0017
   7  e     7   e    2220.5082  2220.5083  -0.0002
   1  e     0   e    2220.5995  2220.5994   0.0001
   8  e     8   e    2220.6904  2220.6903   0.0001
   9  e     9   e    2220.7812  2220.7810   0.0002
  11  e    11   e    2220.8713  2220.8714  -0.0001
*********************************************
```

Table A-4. Observed and calculated transitions in the C-O stretching fundamental band of OC-DCCD dimer (units of 1/cm).

```
*************************************************
  J' symm   J" symm    Obs.       Calc.     Obs-Calc
*************************************************
  11   e     12   e    2148.4928  2148.4930  -0.0002
   9   e     10   e    2148.6760  2148.6762  -0.0002
   8   e      9   e    2148.7673  2148.7674  -0.0001
   7   e      8   e    2148.8583  2148.8584  -0.0001
   6   e      7   e    2148.9490  2148.9491  -0.0001
   5   e      6   e    2149.0394  2149.0395  -0.0001
   4   e      5   e    2149.1296  2149.1296   0.0000
   3   e      4   e    2149.2195  2149.2195  -0.0000
   2   e      3   e    2149.3091  2149.3091   0.0000
   1   e      2   e    2149.3983  2149.3983   0.0000
   0   e      1   e    2149.4873  2149.4873  -0.0000
   1   e      0   e    2149.6644  2149.6644   0.0001
   2   e      1   e    2149.7525  2149.7525   0.0000
   3   e      2   e    2149.8403  2149.8402   0.0001
   4   e      3   e    2149.9277  2149.9277   0.0000
   5   e      4   e    2150.0149  2150.0148   0.0001
   6   e      5   e    2150.1017  2150.1016   0.0001
   7   e      6   e    2150.1882  2150.1882   0.0001
   8   e      7   e    2150.2744  2150.2743   0.0000
   9   e      8   e    2150.3602  2150.3602  -0.0000
  10   e      9   e    2150.4460  2150.4458   0.0002
  11   e     10   e    2150.5312  2150.5310   0.0002
  12   e     11   e    2150.6161  2150.6159   0.0002
  13   e     12   e    2150.7006  2150.7006   0.0001
  14   e     13   e    2150.7847  2150.7849  -0.0001
*************************************************
```

Table A-5. Observed and calculated transitions in nu3+nu9 combination band of OC-DCCD dimer in the region of C-O stretching fundamental (units of 1/cm).

```
**************************************************
 J'  symm   J"  symm    Obs.       Calc.     Obs-Calc
**************************************************
 10   e     11   e    2167.6845  2167.6848  -0.0002
  9   e     10   e    2167.7577  2167.7579  -0.0002
  8   e      9   e    2167.8325  2167.8327  -0.0002
  7   e      8   e    2167.9092  2167.9093  -0.0001
  6   e      7   e    2167.9874  2167.9875  -0.0002
  5   e      6   e    2168.0674  2168.0675  -0.0001
  4   e      5   e    2168.1491  2168.1493  -0.0002
  3   e      4   e    2168.2326  2168.2327  -0.0002
  2   e      3   e    2168.3179  2168.3179  -0.0000
  1   f      2   e    2168.4959  2168.4958   0.0001
  1   f      1   e    2168.5002  2168.5002   0.0001
  2   f      2   e    2168.5068  2168.5067   0.0001
  3   f      3   e    2168.5156  2168.5155   0.0001
  4   f      4   e    2168.5265  2168.5265   0.0001
  5   f      5   e    2168.5397  2168.5396   0.0001
  6   f      6   e    2168.5549  2168.5548   0.0000
  7   f      7   e    2168.5723  2168.5723   0.0001
  1   e      0   e    2168.5841  2168.5840   0.0001
  8   f      8   e    2168.5919  2168.5918   0.0001
  9   f      9   e    2168.6136  2168.6134   0.0001
 11   f     11   e    2168.6371  2168.6372  -0.0001
  2   f      1   e    2168.6631  2168.6630   0.0001
 12   e     12   e    2168.6762  2168.6761   0.0001
  3   f      2   e    2168.6906  2168.6909  -0.0002
  4   e      3   e    2168.7701  2168.7700   0.0001
  5   e      4   e    2168.8657  2168.8656   0.0001
  6   e      5   e    2168.9630  2168.9629   0.0001
  7   e      6   e    2169.0619  2169.0619  -0.0000
  8   e      7   e    2169.1626  2169.1625   0.0000
  9   e      8   e    2169.2649  2169.2648   0.0001
```

| | | | | | | |
|---|---|---|---|---|---|---|
| 10 | e | 9 | e | 2169.3688 | 2169.3688 | 0.0001 |
| 11 | e | 10 | e | 2169.4743 | 2169.4744 | -0.0000 |
| 12 | e | 11 | e | 2169.5817 | 2169.5815 | 0.0001 |
| 13 | e | 12 | e | 2169.6904 | 2169.6902 | 0.0002 |
| 14 | e | 13 | e | 2169.8005 | 2169.8005 | 0.0000 |

**********************************************

Table A-6. Observed and calculated transitions in nu3+nu8 combination band of OC-DCCD dimer in the region of C-O stretching fundamental (units of 1/cm).

```
***************************************************
  J' symm   J" symm    Obs.        Calc.      Obs-Calc
***************************************************
   8  f      8  e     2217.4334  2217.4346  -0.0012  Perturbation allowed transition
   9  e     10  e     2217.4420  2217.4439  -0.0018
   8  e      9  e     2217.4647  2217.4658  -0.0011
  10  e     11  e     2217.4894  2217.4890   0.0005
   7  e      8  e     2217.5192  2217.5191   0.0001
   6  e      7  e     2217.5883  2217.5876   0.0008
   5  e      6  e     2217.6651  2217.6642   0.0009
   9  f      9  e     2217.7092  2217.7068   0.0023  Perturbation allowed transition
   4  e      5  e     2217.7460  2217.7456   0.0005
   3  e      4  e     2217.8299  2217.8298   0.0001
  10  f     10  e     2217.8914  2217.8911   0.0003
   2  e      3  e     2217.9155  2217.9159  -0.0004
  11  f     11  e     2217.9837  2217.9862  -0.0026
   1  e      2  e     2218.0025  2218.0032  -0.0007
  12  f     12  e     2218.0375  2218.0363   0.0011
   1  f      1  e     2218.1810  2218.1811  -0.0002
   2  f      2  e     2218.1828  2218.1837  -0.0008
   3  f      3  e     2218.1876  2218.1880  -0.0004
   4  f      4  e     2218.1949  2218.1950  -0.0001
   5  f      5  e     2218.2062  2218.2061   0.0001
   6  f      6  e     2218.2245  2218.2243   0.0002
   7  f      7  e     2218.2565  2218.2564   0.0001
   1  e      0  e     2218.2685  2218.2692  -0.0008
   8  f      8  e     2218.3192  2218.3190   0.0002
   2  e      1  e     2218.3590  2218.3593  -0.0003
   3  e      2  e     2218.4510  2218.4506   0.0004
   9  f      9  e     2218.4554  2218.4544   0.0010
   4  e      3  e     2218.5447  2218.5436   0.0011
  10  e      9  e     2218.6142  2218.6133   0.0009  Perturbation allowed transition
   5  e      4  e     2218.6410  2218.6396   0.0014
```

| J' | symm | J" | symm | Obs. | Calc. | Obs-Calc | |
|---|---|---|---|---|---|---|---|
| 10 | f | 10 | e | 2218.7237 | 2218.7246 | -0.0009 | Perturbation allowed transition |
| 7 | e | 6 | e | 2218.8497 | 2218.8489 | 0.0008 | |
| 11 | e | 10 | e | 2218.8699 | 2218.8727 | -0.0028 | |
| 8 | e | 7 | e | 2218.9723 | 2218.9728 | -0.0005 | |
| 12 | e | 11 | e | 2219.0624 | 2219.0605 | 0.0019 | |
| 9 | e | 8 | e | 2219.1266 | 2219.1279 | -0.0013 | |
| 10 | e | 9 | e | 2219.3513 | 2219.3500 | 0.0013 | |

**************************************************

Table A-7. Observed and calculated transitions in the C-O stretching fundamental band of OC-DCCH dimer (units of 1/cm).

**************************************************

| J' | symm | J" | symm | Obs. | Calc. | Obs-Calc |
|---|---|---|---|---|---|---|
| 7 | e | 8 | e | 2148.813 | 2148.8133 | -0.0003 |
| 6 | e | 7 | e | 2148.908 | 2148.9084 | -0.0004 |
| 5 | e | 6 | e | 2149.002 | 2149.0032 | -0.0004 |
| 4 | e | 5 | e | 2149.098 | 2149.0978 | 0.0005 |
| 3 | e | 4 | e | 2149.191 | 2149.1920 | -0.0001 |
| 2 | e | 3 | e | 2149.286 | 2149.2860 | 0.0002 |
| 1 | e | 2 | e | 2149.379 | 2149.3796 | 0.0000 |
| 0 | e | 1 | e | 2149.472 | 2149.4730 | -0.0003 |
| 1 | e | 0 | e | 2149.658 | 2149.6587 | 0.0003 |
| 4 | e | 3 | e | 2149.934 | 2149.9347 | 0.0000 |
| 5 | e | 4 | e | 2150.026 | 2150.0261 | -0.0001 |
| 6 | e | 5 | e | 2150.117 | 2150.1172 | 0.0001 |
| 7 | e | 6 | e | 2150.208 | 2150.2079 | 0.0002 |
| 8 | e | 7 | e | 2150.298 | 2150.2984 | 0.0000 |
| 9 | e | 8 | e | 2150.388 | 2150.3885 | 0.0004 |
| 10 | e | 9 | e | 2150.478 | 2150.4783 | 0.0000 |
| 11 | e | 10 | e | 2150.567 | 2150.5679 | -0.0002 |

**************************************************

Table A-8. Observed and calculated transitions in the C-O stretching fundamental band of OC-(C2H2)2 trimer (units of 1/cm).

```
****************************************************
  J' Ka' Kc'    J" Ka" Kc"  Observed  Calculated  Obs-Calc
****************************************************
  8  3  6    9  3  7  2148.3235  2148.3235   0.0000
  7  2  5    8  2  6  2148.3559  2148.3561  -0.0002
  7  2  5    8  2  6  2148.3562  2148.3561   0.0001
  5  2  3    6  2  4  2148.5083  2148.5081   0.0002
  5  3  2    6  3  3  2148.5223  2148.5223   0.0000
  4  2  2    5  2  3  2148.5948  2148.5947   0.0001
  4  2  3    5  2  4  2148.6421  2148.6431  -0.0010
  3  3  1    4  3  2  2148.7094  2148.7085   0.0008
  3  2  2    4  2  3  2148.7154  2148.7162  -0.0008
  3  0  3    4  0  4  2148.7425  2148.7424   0.0002
  2  1  1    3  1  2  2148.7685  2148.7687  -0.0001
  2  0  2    3  0  3  2148.8047  2148.8046   0.0001
  2  1  2    3  1  3  2148.8175  2148.8173   0.0002
  1  1  0    2  1  1  2148.8519  2148.8523  -0.0005
  1  1  1    2  1  2  2148.8861  2148.8851   0.0010
  7  3  5    7  3  4  2148.9366  2148.9363   0.0003
  0  0  0    1  0  1  2148.9473  2148.9463   0.0010
  6  3  4    6  3  3  2148.9806  2148.9814  -0.0008
  1  0  1    0  0  0  2149.1021  2149.1016   0.0005
  2  1  2    1  1  1  2149.1633  2149.1630   0.0003
  2  0  2    1  0  1  2149.1752  2149.1750   0.0002
  2  1  1    1  1  0  2149.1950  2149.1954  -0.0004
  3  1  3    2  1  2  2149.2296  2149.2299  -0.0003
  3  0  3    2  0  2  2149.2419  2149.2418   0.0001
  3  2  1    2  2  0  2149.2711  2149.2714  -0.0003
  3  1  2    2  1  1  2149.2775  2149.2777  -0.0001
  4  1  4    3  1  3  2149.2947  2149.2946   0.0000
  4  0  4    3  0  3  2149.3031  2149.3030   0.0001
  4  2  3    3  2  2  2149.3303  2149.3311  -0.0008
  5  2  4    4  2  3  2149.4020  2149.4022  -0.0001
  6  0  6    5  0  5  2149.4212  2149.4215  -0.0003
```

|  J' Ka' Kc' | J" Ka" Kc" | Observed | Calculated | Obs-Calc |
|---|---|---|---|---|
| 5  1  4 | 4  1  3 | 2149.4261 | 2149.4270 | -0.0009 |
| 5  2  3 | 4  2  2 | 2149.4483 | 2149.4486 | -0.0003 |
| 6  2  5 | 5  2  4 | 2149.4704 | 2149.4704 | -0.0000 |
| 7  1  7 | 6  1  6 | 2149.4812 | 2149.4804 | 0.0008 |
| 7  0  7 | 6  0  6 | 2149.4812 | 2149.4812 | -0.0001 |
| 6  1  5 | 5  1  4 | 2149.4912 | 2149.4911 | 0.0001 |
| 6  3  4 | 5  3  3 | 2149.4970 | 2149.4975 | -0.0004 |
| 6  3  3 | 5  3  2 | 2149.5214 | 2149.5216 | -0.0002 |
| 8  0  8 | 7  0  7 | 2149.5411 | 2149.5412 | -0.0001 |
| 8  1  8 | 7  1  7 | 2149.5411 | 2149.5409 | 0.0003 |
| 7  1  6 | 6  1  5 | 2149.5501 | 2149.5499 | 0.0002 |
| 7  3  5 | 6  3  4 | 2149.5720 | 2149.5720 | 0.0001 |
| 7  4  3 | 6  4  2 | 2149.5917 | 2149.5909 | 0.0008 |
| 9  1  9 | 8  1  8 | 2149.6010 | 2149.6011 | -0.0001 |
| 9  0  9 | 8  0  8 | 2149.6010 | 2149.6012 | -0.0002 |
| 7  3  4 | 6  3  3 | 2149.6134 | 2149.6134 | -0.0001 |
| 8  2  6 | 7  2  5 | 2149.6781 | 2149.6776 | 0.0005 |
| 8  3  5 | 7  3  4 | 2149.7025 | 2149.7025 | 0.0000 |
| 9  4  6 | 8  4  5 | 2149.7394 | 2149.7394 | -0.0000 |
| 9  2  7 | 8  2  6 | 2149.7394 | 2149.7388 | 0.0006 |
| 10 2  8 | 9  2  7 | 2149.7947 | 2149.7952 | -0.0005 |

*****************************************************

Table A-9. Observed and calculated transitions in the C-O stretching fundamental band of OC-(C2D2)2 trimer (units of 1/cm).

*****************************************************

|  J' Ka' Kc' | J" Ka" Kc" | Observed | Calculated | Obs-Calc |
|---|---|---|---|---|
| 7  2  6 | 8  2  7 | 2148.7031 | 2148.7052 | -0.0021 |
| 8  0  8 | 9  0  9 | 2148.7031 | 2148.7030 | 0.0001 |
| 8  1  8 | 9  1  9 | 2148.7031 | 2148.7032 | -0.0001 |
| 6  5  2 | 7  5  3 | 2148.7351 | 2148.7351 | 0.0000 |
| 6  3  4 | 7  3  5 | 2148.7351 | 2148.7350 | 0.0001 |

| | | | | | | | | |
|---|---|---|---|---|---|---|---|---|
| 6 | 1 | 5 | 7 | 1 | 6 | 2148.7521 | 2148.7512 | 0.0008 |
| 7 | 1 | 7 | 8 | 1 | 8 | 2148.7617 | 2148.7615 | 0.0002 |
| 7 | 0 | 7 | 8 | 0 | 8 | 2148.7617 | 2148.7611 | 0.0006 |
| 6 | 1 | 6 | 7 | 1 | 7 | 2148.8194 | 2148.8199 | -0.0006 |
| 6 | 0 | 6 | 7 | 0 | 7 | 2148.8194 | 2148.8189 | 0.0005 |
| 5 | 2 | 4 | 6 | 2 | 5 | 2148.8302 | 2148.8307 | -0.0004 |
| 4 | 2 | 2 | 5 | 2 | 3 | 2148.8522 | 2148.8517 | 0.0004 |
| 4 | 2 | 3 | 5 | 2 | 4 | 2148.8971 | 2148.8968 | 0.0003 |
| 4 | 0 | 4 | 5 | 0 | 5 | 2148.9333 | 2148.9330 | 0.0003 |
| 3 | 0 | 3 | 4 | 0 | 4 | 2148.9896 | 2148.9894 | 0.0001 |
| 3 | 1 | 3 | 4 | 1 | 4 | 2148.9982 | 2148.9983 | -0.0001 |
| 2 | 1 | 1 | 3 | 1 | 2 | 2149.0143 | 2149.0144 | -0.0001 |
| 2 | 2 | 0 | 3 | 2 | 1 | 2149.0230 | 2149.0228 | 0.0002 |
| 2 | 0 | 2 | 3 | 0 | 3 | 2149.0479 | 2149.0479 | 0.0000 |
| 2 | 1 | 2 | 3 | 1 | 3 | 2149.0601 | 2149.0601 | 0.0000 |
| 1 | 1 | 0 | 2 | 1 | 1 | 2149.0929 | 2149.0929 | -0.0001 |
| 1 | 0 | 1 | 2 | 0 | 2 | 2149.1115 | 2149.1115 | 0.0000 |
| 1 | 1 | 1 | 2 | 1 | 2 | 2149.1239 | 2149.1236 | 0.0003 |
| 5 | 2 | 3 | 5 | 2 | 4 | 2149.3397 | 2149.3394 | 0.0003 |
| 3 | 1 | 2 | 3 | 1 | 3 | 2149.3431 | 2149.3435 | -0.0005 |
| 2 | 1 | 2 | 1 | 1 | 1 | 2149.3836 | 2149.3842 | -0.0006 |
| 2 | 1 | 1 | 1 | 1 | 0 | 2149.4143 | 2149.4143 | 0.0001 |
| 3 | 1 | 3 | 2 | 1 | 2 | 2149.4467 | 2149.4470 | -0.0003 |
| 3 | 0 | 3 | 2 | 0 | 2 | 2149.4581 | 2149.4583 | -0.0002 |
| 3 | 2 | 2 | 2 | 2 | 1 | 2149.4726 | 2149.4725 | 0.0001 |
| 4 | 1 | 4 | 3 | 1 | 3 | 2149.5080 | 2149.5079 | 0.0001 |
| 4 | 0 | 4 | 3 | 0 | 3 | 2149.5159 | 2149.5160 | -0.0001 |
| 4 | 2 | 3 | 3 | 2 | 2 | 2149.5420 | 2149.5417 | 0.0003 |
| 5 | 2 | 4 | 4 | 2 | 3 | 2149.6085 | 2149.6086 | -0.0001 |
| 5 | 2 | 3 | 4 | 2 | 2 | 2149.6506 | 2149.6511 | -0.0005 |
| 6 | 2 | 5 | 5 | 2 | 4 | 2149.6726 | 2149.6729 | -0.0003 |
| 7 | 0 | 7 | 6 | 0 | 6 | 2149.6836 | 2149.6840 | -0.0005 |
| 7 | 1 | 7 | 6 | 1 | 6 | 2149.6836 | 2149.6831 | 0.0004 |
| 6 | 1 | 5 | 5 | 1 | 4 | 2149.6927 | 2149.6926 | 0.0001 |
| 6 | 3 | 4 | 5 | 3 | 3 | 2149.6977 | 2149.6976 | 0.0001 |
| 6 | 2 | 4 | 5 | 2 | 3 | 2149.7297 | 2149.7294 | 0.0003 |
| 7 | 2 | 6 | 6 | 2 | 5 | 2149.7349 | 2149.7348 | 0.0001 |

```
 8  1  8     7  1  7   2149.7404  2149.7403   0.0001
 8  0  8     7  0  7   2149.7404  2149.7407  -0.0002
 7  3  5     6  3  4   2149.7680  2149.7676   0.0004
 8  2  6     7  2  5   2149.8674  2149.8673   0.0001
11  1 11    10  1 10   2149.9108  2149.9108   0.0001
11  0 11    10  0 10   2149.9108  2149.9108   0.0000
 9  4  5     8  4  4   2149.9536  2149.9537  -0.0001
*************************************************
```